\documentclass{iaus}
\usepackage{graphicx}

\title[] 
{The locations of SNe Ib/c and their comparison to those of WR stars and GRBs}

\author[Giorgos Leloudas]   
{Giorgos Leloudas$^{1,2}$
 }

\affiliation{$^1$The Oskar Klein Centre, Department of Physics, Stockholm University, \\ 106 91 Stockholm, Sweden \\ email: {\tt giorgos.leloudas@fysik.su.se} \\[\affilskip]
$^2$Dark Cosmology Centre, Niels Bohr Institute, University of Copenhagen, \\ 2100 Copenhagen, Denmark}

\pubyear{2012}
\volume{279}  
\jname{Death of Massive Stars: Supernovae and Gamma-Ray Bursts}
\editors{P. Roming, N. Kawai \& E. Pian, eds.}
\begin{document}

\maketitle

\begin{abstract}

The locations of long GRBs and stripped supernovae are compared to those of their favored progenitors, WR stars, and their sub-classes.
Compared to  \cite[{{Leloudas} {et~al.} (2010)}]{Leloudas2010}, we have doubled the number of galaxies with suitable WR data. 
In the combined sample, WC stars are found, on average, in brighter locations than WN stars. 
The WN distribution is fully consistent with the one of SNe~Ib, 
while it is inconsistent with those of SNe~II, Ic and GRBs. 
The WC distribution is both consistent with SNe~Ib and Ic. It is inconsistent with SNe~II, and marginally consistent with GRBs.
Furthermore, we present a spectroscopic study of the locations of SNe~Ib/c.
The average metallicity in the environments of  SNe~Ic  is found to be a little higher than for SNe~Ib, but the difference is small and not significant within our sample.
Under the assumption that the SN regions were formed in an instantaneous burst of star formation, 
we find that a fraction of them appear older than what is allowed in order to host SNe~Ib/c from single massive stars.
Within this framework, these SNe must come from lower mass binaries.

\end{abstract}

\firstsection 
\section{Introduction: stripped supernovae and open questions}

Type Ib/c supernovae (SNe) are among the few astronomical objects that do not show H in their spectra. 
It is therefore believed that they are the explosions of stars that have shed their outer H envelope (SNe~Ib),
or even (most of) their He envelope (SNe~Ic).
For this reason, they are known as \textit{stripped} SNe.
To date, there has been no direct progenitor detection of a SN~Ib/c in pre-explosion images (\cite[{{Smartt} 2009}]{smarttARAA}).
Nevertheless, it is widely believed that stripped SNe are the explosions of Wolf-Rayet (WR) stars, 
exactly because these stars appear to be H-free.
Depending on their spectroscopic appearance, WR stars can be broadly divided into nitrogen-rich (WN) and carbon-rich (WC) stars.
A comprehensive review of WR stars, their sub-classes, and their properties is given by \cite[{{Crowther} (2007)}]{CrowtherReview}. 
WN stars are expected to die as SNe~Ib, while WC stars as SNe~Ic, although variations to this general picture might exist, depending on the
progenitor metallicity and initial mass (e.g. \cite[{{Georgy} {et~al.} 2009}]{Georgy2009}).
Even if the connection of SNe~Ib/c to WR stars seems like the only plausible scenario, it is important to stress that it awaits confirmation,
especially in its details.

Another fundamental question concerning stripped SNe and their progenitors is the \textit{way} that they lose their H envelopes.
Massive stars lose mass through stellar winds and this is one of the main theoretical channels leading to a SN~Ib/c.
For single stars to reach the WR phase it is estimated that they need to start their lives with an initial mass $>$~25~M$_{\odot}$ (\cite[{{Crowther} 2007}]{CrowtherReview}).
Another way for a star to lose its H envelope, is through evolution in a binary system (e.g. \cite[{{Podsiadlowski} {et~al.} 1992}]{Podsial92}).
In this scenario, the outer layers of the primary star are expelled through interaction with its binary companion (Roche-lobe overflow or common envelope evolution).
It is important to note that through the binary channel, it is possible to obtain stripped stars from initial masses substantially lower than from the single progenitor channel.

Direct searches for SN progenitors have not yielded any results yet and, even in the best case, are not expected to give more than a handful of detections during the next decade.
Studies of the SN properties themselves (light-curves, spectra, etc) are informative, but they suffer from small numbers and the difficulty of acquiring suitable homogeneous datasets. 
Another powerful means of probing the nature of these explosions is by studying their environments. 
Although indirect, `environmental' methods have statistical power (as large samples can be constructed) and can be used in complementary manner to direct methods in order to attack the same problem from different angles. 
Here, we concentrate on some of these methods as they have been applied to stripped SNe: in Section~\ref{sec:WR} we  compare the locations of SNe~Ib/c (and long GRBs) within their host galaxy with those of WR stars, while in Section~\ref{sec:spec} we  describe direct spectroscopic observations of the birthplaces of SNe~Ib/c, with the purpose of studying their metallicity and stellar population ages.

\section{Comparison with locations of WR stars}
\label{sec:WR}

The main driver motivating this comparison is the following: 
\textit{if WR stars are the progenitors of stripped SNe and long GRBs, they must also be found in similar locations within their hosts}.
The distribution of long GRBs with respect to their host galaxy light was studied by \cite[{{Fruchter} {et~al.} (2006)}]{F06}, 
while those of different SN sub-types by \cite[{{Kelly} {et~al.} (2008)}]{K08}.
The statistical diagnostic used is the \textit{fractional flux}, which describes the fraction of light in host locations 
fainter than the explosion locations, over the integrated galaxy light. 
Simply put, a fractional flux value of 0 means that the explosion took place in a location without flux, 
while a value of 1 means that the explosion took place in the brightest location of the host.
A similar statistic was used by \cite[{{Anderson} \& {James} (2008)}]{Anderson2008}, while 
other authors (e.g. \cite[{{Larsson} {et~al.} 2007}]{Larsson2007}) have used this notion in theoretical studies.
\cite[{{Fruchter} {et~al.} (2006)}]{F06} showed that GRBs showed a strong preference for occurring in the brightest locations of their host.
\cite[{{Kelly} {et~al.} (2008)}]{K08}, studying a more nearby sample of SN hosts drawn from SDSS, showed that 
SNe~II, Ib, and Ic occur in progressively brighter locations, with the distribution of SNe~Ic being broadly consistent with the one of GRBs.

In  \cite[{{Leloudas} {et~al.} (2010)}]{Leloudas2010}, hereafter L10, we extended this analysis to compare with the locations of WR stars within their host galaxies.
For this purpose, we used the work of Crowther and collaborators, who in the recent years have mapped the population of WR stars in a number of nearby galaxies. 
What makes these papers very attractive for this kind of study is that a detailed estimate of the \textit{completion level} of the surveys is given. 
We have excluded from our study many galaxies whose information on the WR population is known, but suffer from some kind of incompleteness.
At the time of publication of L10, this analysis was possible for two galaxies: M~83 
(\cite[{{Hadfield} {et~al.} 2005}]{M83}) and NGC~1313  (\cite[{{Hadfield} \& {Crowther} 2007}]{NGC1313}).
At present, suitable data are available for two more galaxies: NGC~7793 (\cite[{{Bibby} \& {Crowther} 2010}]{Bibby2010}) and  NGC~5068 (\cite[{{Bibby} \& {Crowther}  2012}]{Bibby2012}). In this proceeding, we extend our analysis to include these new data. It should thus be regarded as a continuation and complementary to the original paper (L10).
  
A conceptual difference between our approach and the one of \cite[{{Fruchter} {et~al.} (2006)}]{F06} and \cite[{{Kelly} {et~al.} (2008)}]{K08}, is that 
these studies examine many host galaxies with one explosion per galaxy, while we only examine a few galaxies with many potential explosion progenitors.
It is therefore important to take into account the general properties of these individual galaxies during the comparison, to make sure that they are not radically different.
Metallicity is a particularly important factor, as both the total number of WR stars and the ratio of WC to WN stars strongly depend on it.
In L10, we showed that M~83 (a galaxy with super-solar metallicity) is a very typical representative of the SN~Ib/c host sample, while NGC~1313 is both fainter and more metal poor.
The two new galaxies have somewhat intermediate properties between these two. As discussed by \cite[{{Bibby} \& {Crowther}  (2010, 2012)}]{Bibby2012}, they both demonstrate important metallicity gradients encompassing both solar and LMC values.
Another important difference is that these galaxies are more nearby.
To enable the comparison, it is necessary to resample them and make them appear as if they were at the median redshift of the SN host sample (L10).

\begin{figure}[b]
\vspace*{-0.5 cm}
\begin{center}
 \includegraphics[width=\textwidth]{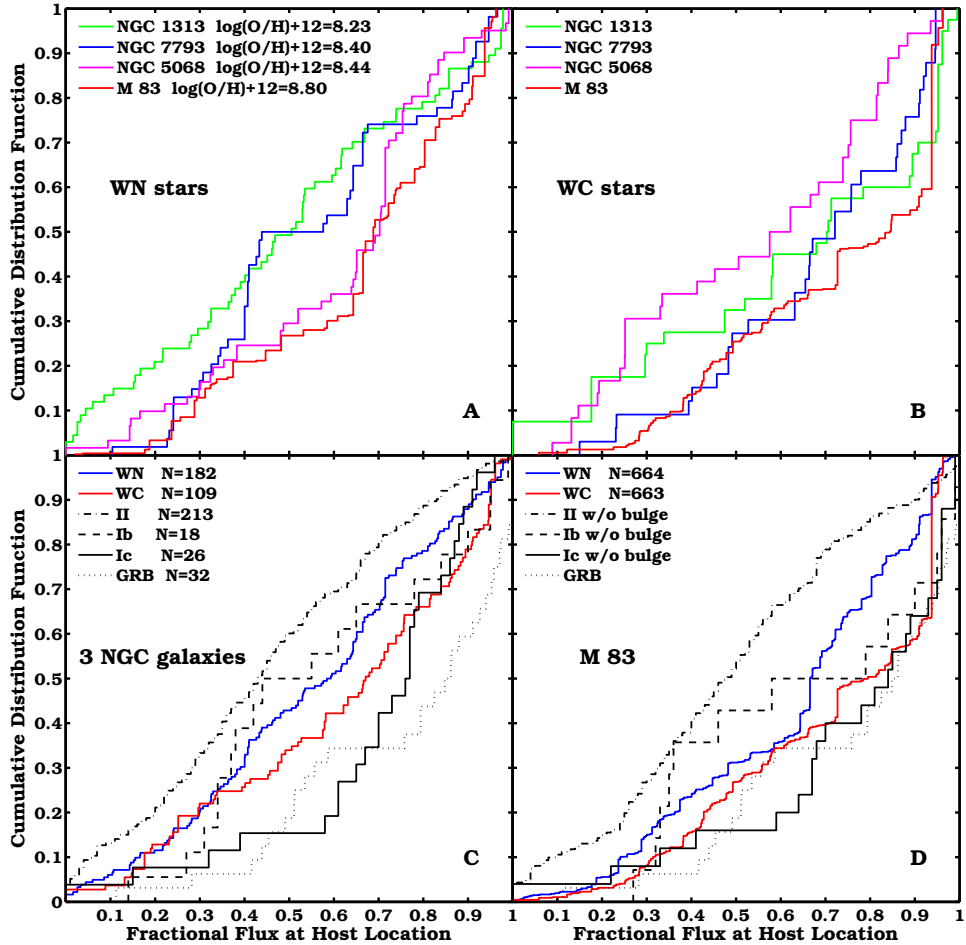} 
\vspace*{-0.5 cm}
 \caption{
Different cumulative distributions of fractional fluxes for WR sub-classes, SNe and GRBs.
Panel A shows the distributions of WN stars in 4 galaxies. 
The average metallicities of these galaxies are also indicated, showing that there is perhaps a metallicity effect.
Panel B shows the same but for WC stars. With the exception of NGC~5068, all WC distributions
are pushed towards brighter locations in their hosts than WN stars.
In panel C, the WN and WC distributions of NGC~1313, NGC~7793 and NGC~5068 have been combined
and are compared to the SN and GRB distributions of   \cite[{{Kelly} {et~al.} (2008)}]{K08} and \cite[{{Fruchter} {et~al.} (2006)}]{F06}.
WC stars are on average found in brighter locations than WN stars.
M~83 that is quite different from the other 3 galaxies in terms of metallicity and number of WR stars, is shown separately in panel D.
In this panel, the distributions are shown after the removal of the bulge light  (\cite[{{Kelly} {et~al.} 2008}]{K08}).
 }
   \label{fig1}
\end{center}
\end{figure}

Figure~\ref{fig1}  contains a comparison of the WN and the WC distributions of these 4 galaxies (panels A and B). 
The WN distributions  are not identical: in the case of NGC~1313, it almost follows the host galaxy light,
while the others move progressively toward brighter locations.
As a matter of fact, this appears to be happening  as a function of host metallicity.
At this stage, we just point this out as an interesting finding, although it might disappear with a larger sample, and it is not straightforward to explain. 
Our favorite explanation is that at the low metal content of NGC~1313, it is more difficult for WR stars to lose all their H, and that WN stars might retain enough H to appear spectroscopically as a SN~II (\cite[{{Georgy} {et~al.} 2009}]{Georgy2009}).
In the case of the WC distributions, the one that stands out is NGC~5068.
This is the only galaxy in this sample, for which the WC distribution is not skewed to brighter pixels than the WN, as also pointed out by \cite[{{Bibby} \& {Crowther}  (2012)}]{Bibby2012}.
One of the main conclusions of L10 was that 
WC stars are found on average in brighter locations (and therefore closer to star-forming regions) of their hosts than WN stars.
This statement is therefore now true for 3 out of 4 galaxies. It will be interesting to see how this will evolve in the future with a larger sample.

It is possible to combine the distributions from the different galaxies, although this approach has  weaknesses because all distributions  have particularities (both physical and related to the observation methods).
Especially M~83 stands out from the rest of the sample: 
(i) it has a super-solar metallicity that is very different from the average metallicities of the other 3 galaxies (that ignoring metallicity gradients are not that different); 
(ii) it has a WR population far greater than all the others added together (and therefore any combined sample would be heavily weighted towards M~83).
For this reason, it was decided to only combine the distributions from the other 3 galaxies. These are presented in  panel C of Fig.~\ref{fig1}, while M~83 is shown separately in panel D.  
In addition, for M~83, comparisons are made after the removal of the bulge light (\cite[{{Kelly} {et~al.} 2008}]{K08}), since the WR survey is incomplete in this region of the galaxy (\cite[{{Hadfield} {et~al.} 2005}]{M83}).  

The results for M~83 were discussed in  L10. Briefly, WC stars are found (on average) in brighter locations than WN stars and both distributions are statistically compatible with SNe~Ib/c. They are incompatible with SNe~II, while WC stars are marginally compatible with GRBs.
In the combined sample, which now contains over 100 WN and WC stars including the most secure photometric candidates,  
we observe again that the average WC distribution appears in brighter pixels than the average WN distribution.
A KS test between the two yields $p = 4.2$\%. 
The WN distribution is highly compatible with SNe~Ib ($p = 98.4$\%) while it is highly incompatible with SNe~II, SNe~Ic and GRBs (p-values of 0.1\%, 0.7\% and 0\% respectively).
The WC distribution is again mostly compatible with SNe~Ib ($p = 37.0$\%), but also with SNe~Ic ($p = 12.3$\%).
It is marginally compatible with GRBs ($p = 1.9$\%), while the association with SNe~II is excluded ($p = 0$\%).
We believe that these results are in nice agreement with the theoretical expectations and with the idea that WC stars originate from more massive progenitors than WN stars.

WN stars sometimes evolve to become WC stars, depending on the initial mass and metallicity. In a Monte Carlo simulation we allowed a fraction of WN stars to be included to the WC fractional flux distribution. For the combined sample, we estimate that this fraction is of the order of 20\% (see L10 for details). The simulation showed that this effect does not alter our general conclusions, although the specific p-values change. Similar simulations were executed by  L10 for M~83 as well as for the effect of photometric candidates and the WR number errors.


\section{Direct spectroscopic observations}
\label{sec:spec}

There are a number of predictions that are related to the single progenitor scenario, and that could be used to test it. 
First, mass stripping must be the result of stellar winds. Stellar winds are driven by metals and their strength increases with metallicity
(e.g. \cite[{{Vink} \& {de Koter} 2005}]{vink2005}).
It is therefore anticipated that SNe~Ic will be found, on average, in more metal-rich environments than SNe~Ib, if they result from massive stars.
Second, massive stars have short lifetimes. Therefore, if SNe~Ib/c come from single massive stars, they must be found close to areas of recent star formation.

{\underline{\it Metallicities}}. The role of metallicity at the locations of SNe, has been examined by many authors.
Until recently, however, most studies were either restricted to special events (\cite[{{Sollerman} {et~al.} 2005}]{sollerman05}, \cite[{{Modjaz} {et~al.} 2008}]{modjaz2008}), or were based on abundances measured at the host galaxy nucleus
(e.g. \cite[{{Prieto} {et~al.} 2008}]{Prieto2008}) or on proxies of metallicity (\cite[{{Boissier} \& {Prantzos} 2009}]{BP09}, \cite[{{Anderson} \& {James} 2009}]{Anderson2009}). Direct metallicity measurements at the locations of normal stripped SNe were presented by \cite[{{Anderson} {et~al.} (2010)}]{andersonCCmetal}, \cite[{{Modjaz} {et~al.} (2011)}]{modjazIbcMetal} and   \cite[{{Leloudas} {et~al.} (2011)}]{Leloudas2011}.

Within our sample, we find an average metallicity of 8.52$\pm$0.05 dex for SNe~Ib and 8.60$\pm$0.08 dex for SNe~Ic (\cite[{{Leloudas} {et~al.} 2011}]{Leloudas2011}). These numbers are given in the N2 scale of  \cite[{{Pettini} \& {Pagel} (2004)}]{PP04} and  include the systematic uncertainty related to the calibration of this relation. 
The contamination from stellar continuum was removed by fitting \cite[{{Bruzual} \& {Charlot} (2003)}]{BC03} models, but this procedure did not significantly affect our results. We conclude that there might be hints that SNe~Ic are found in more metal-rich environments than SNe~Ib, but this difference is far from being statistically significant. A KS test between the two distributions gives the non-decisive p-value of 17\%.

The other two studies did not yield consistent results: \cite[{{Anderson} {et~al.} (2010)}]{andersonCCmetal} find almost equal metallicities between the two sub-classes, but  \cite[{{Modjaz} {et~al.} (2011)}]{modjazIbcMetal} report on a difference of 0.20 dex and a KS test p-value of 1\%.
This might be partly due to the different biases affecting the different studies.
Our sample, unlike the one by   \cite[{{Anderson} {et~al.} (2010)}]{andersonCCmetal} and similar to the one of \cite[{{Modjaz} {et~al.} (2011)}]{modjazIbcMetal},
 comprises SNe in both targeted and non-targeted surveys (in almost equal ratios) and there is evidence that the latter are found in lower metallicity than the former (\cite[{{Leloudas} {et~al.} 2011}]{Leloudas2011}). 

Even in the last two studies, however, it is true that this bias is not properly quantified.
In addition, the difference between SNe~Ib and Ic, if it exists, appears to be small and comparable to the systematic errors affecting the metallicity calibrators.
For this reason, if a significant difference is to be sought, a large and well-controlled sample will be needed (see also talk by M.~Modjaz; this volume).
In any case, we think that it is unlikely that this difference can yield conclusive answers on the issue of binarity of stripped SNe.

\begin{figure}[b]
\begin{center}
 \includegraphics[width=\textwidth]{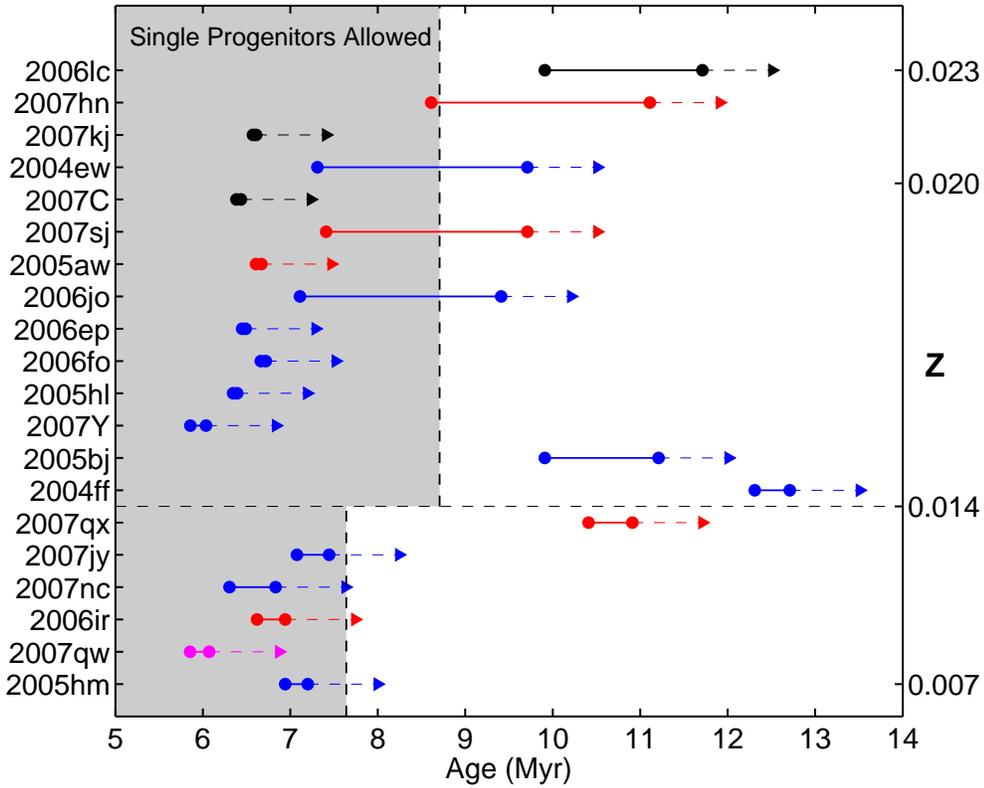} 
 \caption{
Ages of the youngest stellar populations at SN~Ib/c  locations (solid lines), 
estimated through comparing the measured H$\alpha$~EW with the predictions of Starburst99. 
These set a lower limit (dashed arrow) to the age of the SN birthplace.
The SN environments  were ordered by ascending metallicity (some key values are indicated on the righthand axis).
The SNe are color coded according to their type: SNe~Ib in blue, SNe~Ic in red and intermediate SNe~Ib/c in black (an interloper SN~Ia is colored in magenta). 
The horizontal dashed line separates the host regions into two metallicity groups, depending on whether a comparison with 
$Z=0.008$ or $Z=0.02$  models  is more appropriate. 
The vertical dashed lines denote the predictions of the Geneva evolutionary codes for the lifetimes of single massive stars of 25 $M_{\odot}$ at $Z=0.02$ and 30 $M_{\odot}$ at $Z=0.008$. 
The allowed ages for single progenitors of SNe~Ib/c are within the gray-shaded area.
The more massive SNe~Ic, however, have lifetimes $<5$ Myr (i.e. outside this graph).
 }
   \label{fig0}
\end{center}
\end{figure}

{\underline{\it Stellar ages}}. An age estimator  sensitive to the most recent star formation is the equivalent width (EW) of H$\alpha$  (e.g. \cite[{{Leitherer} {et~al.} 1999}]{Starburst99}).  It is thus possible, using our direct spectroscopic observations (\cite[{{Leloudas} {et~al.} 2011}]{Leloudas2011}), to constrain the ages of the local stellar populations in the vicinity of the SN explosions.
It is not important for our purposes that this method is only sensitive to the youngest, ionizing, stellar populations probed by our slit, exactly because what we want  is to place a \textit{lower limit} on the ages of the stars at the SN locations.
The question we want to address is whether there are any SN regions where all stars are older than what is
allowed by evolutionary models for single massive stars.  In that case, by exclusion, the progenitors should be binary systems.

In order to estimate the ages of the young stellar populations, we used the measured H$\alpha$~EW  and compared with the predictions of Starburst99   (\cite[{{Leitherer} {et~al.} 1999}]{Starburst99}) \textit{for instantaneous star formation} (see discussion concerning this limitation below). 
Our metallicity measurements were used  to select the appropriate table. 
The ranges we derived (Fig.~\ref{fig1}) are conservative and include the measurement uncertainty, and the range of corresponding possible ages and IMFs given by \cite[{{Leitherer} {et~al.} (1999)}]{Starburst99}.

These lower limits were then compared with the lifetimes of massive stars, as predicted by the Geneva evolutionary models
(\cite[{{Meynet} \& {Maeder} 2003, 2005}]{Geneva2003}, \cite[{{Georgy} {et~al.} 2009}]{Georgy2009}).
According to these authors, at low metallicity ($Z=0.008$), the lower mass limit above which single stars explode as SNe~Ib/c is 30 $M_{\odot}$, while at solar values ($Z=0.02$) it becomes 25 $M_{\odot}$. The corresponding lifetimes for these limiting stars are $\sim$7.6
and 8.7 Myr, respectively. 
These ages represent  \textit{upper limits} because more massive stars will explode even  sooner.
Consequently, single massive progenitors of SNe~Ib/c can only be found in the gray shaded area of Fig.~\ref{fig1}.
The constraints for He-poor SNe~Ic are even stricter, as they should result from even more massive stars ($>$ 39 $M_{\odot}$;  \cite[{{Georgy} {et~al.} 2009}]{Georgy2009}) and have even shorter lifetimes. As a matter of fact, in Fig.~\ref{fig1} there is no allowed area for SNe~Ic at all.
Therefore, at least four SNe in Fig.~\ref{fig1} seem to have age constraints that are  incompatible with the single progenitor channel, and the number is doubled if we consider the last argument concerning SNe~Ic.
By exclusion, an important fraction  (20-35\%) of the SNe examined should have lower-mass binary progenitors (without this possibility being excluded for the other SNe).

There is one main caveat in our considerations and this is that we have only considered the case of instantaneous star formation.
If the SN birthplace  is still forming stars, no such strict limits can be placed through the H$\alpha$ emission.
We have no means of assessing  the validity of such an assumption for the individual cases and, for this reason, 
it is not possible to unambiguously rule out single progenitors for these `discrepant' SNe.
Nevertheless, we can propose them as good candidates for binarity and 
suggest that this possibility is also investigated by other complementary means.
In addition, it will be interesting to study if the fraction of these discrepancies persists in even larger samples.
To our knowledge, this is the first (and remains the only) attempt to statistically constrain the nature of SN~Ib/c progenitors with this method.

\vspace*{0.5 cm} 
\textbf{Acknowledgments:} I am grateful to Joanne Bibby for providing me with pre-processed images and files for NGC~7793 and NGC~5068, and to 
Max Stritzinger for providing comments on the manuscript.
This work is supported by  the Swedish Research Council through grant No.~623-2011-7117.



\begin{discussion}

\discuss{Paul Crowther}{I have two comments: (1) Extragalactic giant HII regions 
have an age spread of 10+ Myr and are strongly biased towards the youngest (most massive) episode of star formation.
(2) Our extragalactic WR surveys are only sensitive to high mass WR stars while the probable progenitors of many/most SNe Ib/c (He star primary and bright BSG secondary) cannot be detected since the WR wind is swamped by its companion. Not even in the Milky Way.}

\discuss{Giorgos Leloudas}{Indeed, it is true that studying unresolved regions can be problematic. That is why in our paper we did not claim that we can accurately pinpoint the age of the SN progenitor, but only placed a lower limit, based exactly on the signatures of the youngest stars in the region. 
But this is ok, since we wanted to compare with upper limits of single stars lifetimes. It is true that we can probably not do better than this, but, even like this, we discovered some potential discrepancies. 
Concerning the incompleteness of the surveys, I think this is of course important to take under consideration. But we have to try to do as good as we can with the information we have.
}

\discuss{Nathan Smith}{You showed that WN stars and Type II SNe have roughly the same distribution in their host galaxies. This is interesting, because we have strong evidence that most SNe IIP come from lower masses (8-17 M$_{\odot}$) while WN-type Wolf-Rayet stars come from much higher initial masses. This argues strongly against the interpretation that the different locations of different SN types within their host galaxies are due to different initial masses. While you did not make this claim in your talk, it is widely assumed to be the case so it is important to point out that this interpretation is probably wrong.  
It appears that some systematic effect other than the initial mass determines the different distributions of SN types within their host galaxies.}

\discuss{Giorgos Leloudas}{One thing that is important to note, is that this observation is only true for galaxies of lower metallicity, and only statistically significant for NGC~1313. 
As I said, perhaps this is not that surprising, because there have been arguments (by the Geneva group) that at low metallicity WN stars might actually explode as SNe of Type~II, because they retain enough hydrogen to be observed as such.
In addition, the `red supergiant problem' (the missing higher mass SN~II progenitors) indicates exactly that we are missing some part of the picture concerning initial masses. 
I think that we should take the observational evidence and try to explain it, rather than simply using it to dismiss some assumptions that seem logical. 
Indeed, I did not make this claim (initial mass follows light distribution) directly in my talk, as this was not important for my purposes. I restricted myself in comparing locations of explosions and potential progenitors and investigating their compatibility.
However, I have no doubt that there is some kind of relation between the two. And even if it is not a strict correlation in the case of broad-band blue light, it is certainly more informative in the case of H$\alpha$ (see, for example, Joe Anderson's talk).
}

\end{discussion}

\end{document}